\documentstyle[preprint,aps,axodraw]{revtex}
\tightenlines

\newcommand{\reseteqnum}{\setcounter{equation}{0}}
\newcommand{\nn}{\nonumber}

\newcommand{\ovl}[1]{\overline{#1}}
\newcommand{\wt}[1]{\widetilde{#1}}
\newcommand{\eqn}[1]{(\ref{#1})}

\newcommand{\leftD}{{\mathop{D}\limits^{\leftarrow}}}
\newcommand{\rightD}{{\mathop{D}\limits^{\rightarrow}}}
\newcommand{\pslash}{p\kern-1ex /}
\newcommand{\lslash}{l\kern-1ex /}
\newcommand{\Dslash}{{\cal D}\kern-1.5ex /}
\newcommand{\bpsi}{\overline{\psi}}

\begin{document}

\makeatletter
\def\setcaption#1{\def\@captype{#1}}
\makeatother
%

\title{
\vspace{-3.0cm}
\begin{flushright}  
{\normalsize UTCCP-P-40}\\
{\normalsize UTHEP-384}\\
\end{flushright}
Perturbative calculation of improvement coefficients 
        to ${\cal O}(g^2a)$ 
        for bilinear quark operators in lattice QCD} 

\author{Yusuke Taniguchi$^{(a)}$ and Akira Ukawa$^{(a,b)}$}

\address{Institute of Physics$^{(a)}$, University of Tsukuba, 
Tsukuba, Ibaraki-305, Japan \\
Center for Computational Physics$^{(b)}$, University of Tsukuba,
Tsukuba, Ibaraki-305, Japan \\
}

\date{\today}

\maketitle

\begin{abstract}
We calculate the ${\cal O}(g^2 a)$ mixing coefficients of bilinear quark 
operators in lattice QCD using a standard perturbative evaluation of 
on-shell Green's functions.
Our results for the plaquette gluon action are in agreement with those 
previously obtained with the Schr\"odinger functional method. 
The coefficients are also calculated for a class of improved gluon 
actions having six-link terms. 
\end{abstract}


\narrowtext

\section{Introduction}

Symanzik's improvement program\cite{Symanzik83} applied to 
on-shell quantities\cite{LuescherWeisz85} 
attempts to eliminate cut-off dependence order by order by 
an expansion in powers of the lattice spacing $a$.  
To ${\cal O}(a)$ in lattice QCD, this
requires adding the ${\cal O}(a)$ ``clover'' term to the Wilson quark 
action\cite{SheikholeslamiWohlert85}. 
Quark operators also have to be modified by ${\cal O}(a)$ 
counter terms, which generally involve new operators of higher 
dimension\cite{Heatlie91,Jansen96,Luescher9605}.  
In perturbation theory, the tree-level value of the clover coefficient 
and those of the counter terms of quark operators can be easily determined.  
They are sufficient to remove terms of ${\cal O}(g^2a\log a)$ in on-shell 
Green's functions evaluated at one-loop order, as has been explicitly 
demonstrated in Ref.~\cite{Heatlie91}.  
To remove ${\cal O}(g^2a)$ terms which still 
remains, the counter term coefficients for quark operators have to 
be corrected by ${\cal O}(g^2a)$ terms.  For bilinear quark operators, 
these coefficients have been calculated in 
Refs.~\cite{Luescher9606,Sint-Weisz} using the Schr\"odinger functional
technique. 

In this article we analyze the ${\cal O}(g^2a)$ coefficients of bilinear 
quark operators through a standard perturbative treatment of on-shell 
Green's functions of the operators.  One-loop amplitudes  
with external quarks on the mass shell are expanded in powers of $qa$ 
and $m_Ra$,  with $q$ the external momenta and $m_R$ the renormalized 
quark mass, 
which leads to an alternative determination of the coefficients.  
Applying the procedure for the standard
plaquette gluon action, we obtain results which are in agreement with 
those of Refs.~\cite{Luescher9606,Sint-Weisz}.

Another application of our procedure is a calculation of the 
coefficients for gluon actions improved by an addition 
of six-link loop terms to the plaquette action.  We treat three cases:
the action\cite{Weisz83} which is tree-level improved in Symanzik's 
sense to ${\cal O}(a^4)$, and two types of actions\cite{Wilson80,Iwasaki83} 
improved by a renormalization-group treatment.  The results should be 
useful in simulations employing improved actions for gluons. 

This paper is organized as follows.  In Sec.~II we define bilinear quark 
operators examined in this article, and write down renormalization relations
between the renormalized and bare operators.  Analysis of one-loop amplitudes
are carried out in Sec.~III.  Numerical results for the coefficients and 
a comparison with previous work are made in Sec.~IV.  We close with some
concluding remarks in Sec.~V.

\section{Clover quark action and bilinear quark operators}

Consider the clover quark action defined by
\begin{eqnarray}
S_{\rm quark} &=&
a^3 \sum_n \frac{1}{2} \sum_\mu \left(
\bpsi_n (-r + \gamma_\mu) U_{n, \mu} \psi_{n+\hat{\mu}}
+
\bpsi_n (-r - \gamma_\mu) U^\dagger_{n-\mu, \mu}
\psi_{n-\hat{\mu}} \right)
+ (a m_0+4r) \bpsi_n \psi_n 
\nn\\
&-& c_{\rm SW} a^3 \sum_n \sum_{\mu, \nu}
ig \frac{r}{4} \bpsi_n \sigma_{\mu \nu} P_{\mu \nu} (n) \psi_n .
\end{eqnarray}
We wish to construct a renormalized bilinear quark operator of form
\begin{equation}
{\cal O}_R^\Gamma = \left(\bpsi_c \Gamma \psi_c\right)_R,\qquad
\Gamma= 1, \gamma_5, \gamma_\mu, \gamma_\mu \gamma_5, \sigma_{\mu \nu}
\label{eq:bilinear}
\end{equation}
which is improved to $O(a)$ to one-loop order of perturbation 
theory, {\it i.e.,} on-shell matrix elements of the operator have 
no errors of ${\cal O}(a)$, ${\cal O}(g^2 a\log a)$ or ${\cal O}(g^2 a)$
when $a\to 0$ with external momenta and the renormalized quark mass $m_R$
kept fixed. 

Our starting point is the tree-level improved operator of form, 
\begin{eqnarray}
{\cal O} = \bpsi_c \Gamma \psi_c , 
\end{eqnarray}
where the rotated quark fields $\psi_c$ and $\bpsi_c$ are given by 
\begin{eqnarray}
\psi_c &=& \left[ 1-\frac{ar}{4}
 \left( \gamma_\mu \rightD_\mu - m_0 \right) \right] \psi ,
\\
\bpsi_c &=& \bpsi \left[ 1-\frac{ar}{4}
 \left( -\gamma_\mu \leftD_\mu - m_0 \right) \right]
\end{eqnarray}
with the covariant derivative defined as
\begin{eqnarray}
a \rightD_\mu \psi (x) &=&
\frac{1}{2} \left[
 U_\mu (x) \psi (x+\hat{\mu}) - U_\mu^\dagger (x-\hat{\mu})
 \psi (x-\hat{\mu}) \right] ,
\\
a \bpsi (x) \leftD_\mu &=&
\frac{1}{2} \left[
\bpsi (x+\hat{\mu}) U_\mu^\dagger (x)-\bpsi (x-\hat{\mu})
 U_\mu (x-\hat{\mu}) \right].
\end{eqnarray}

It has been demonstrated in Ref.~\cite{Heatlie91} that 
this operator is on-shell improved to ${\cal O}(a)$ and 
${\cal O}(g^2 a \log a)$ with the tree-level value of the 
clover coefficient $c_{SW}=1$. 
It has been noted furthermore that the field rotation can be 
generalized by using the equation of motion $(\Dslash + m_0) \psi =0$ to
\begin{eqnarray}
\psi_c &=& \left[ 1-\frac{ar}{2}
 \left( z \gamma_\mu \rightD_\mu - (1-z) m_0 \right) \right] \psi ,
\label{eqn:rotate-r}
\\
\bpsi_c &=& \bpsi \left[ 1-\frac{ar}{2}
 \left( -z \gamma_\mu \leftD_\mu - (1-z) m_0 \right) \right],
\label{eqn:rotate-l}
\end{eqnarray}
where $z$ is a parameter. 
We then consider a generalized operator given by
\begin{eqnarray}
{\cal O}_0^\Gamma
= \left[ 1+ar \left( 1-z \right) m_0 \right] \bpsi \Gamma \psi
+ z \Gamma^\otimes
- z^2 \Gamma ',
\label{eqn:improved-onshell}
\end{eqnarray}
where $\Gamma^\otimes$ and $\Gamma '$ are
${\cal O}(a)$ and ${\cal O}(a^2)$ vertices defined as
\begin{eqnarray}
\Gamma^\otimes
&=& \frac{ar}{2}
\left( \gamma_\mu \leftD_\mu \Gamma - \Gamma \gamma_\mu \rightD_\mu \right) ,
\\
\Gamma '
&=& \frac{a^2 r^2}{4}
\gamma_\nu \leftD_\nu \Gamma \gamma_\mu \rightD_\mu.
\end{eqnarray}

The one-loop relation expected between the bare operator 
(\ref{eqn:improved-onshell}) and the renormalized improved operator 
(\ref{eq:bilinear}) has the form 
\begin{eqnarray}
{\cal O}^\Gamma_{0} =
Z_\Gamma^{-1} {\cal O}^\Gamma_{\rm R}
- g^2 C_F a m_R\, B_\Gamma\, {\cal O}^\Gamma_{\rm R}
- g^2 C_F a\, C_\Gamma\, \wt{{\cal O}}^{\Gamma}_{\rm R} ,
\label{eqn:renormalization}
\end{eqnarray}
where $C_F$ denotes the second-order Casimir eigenvalue for the quark filed, 
and the last two terms are needed to remove ${\cal O}(g^2 a)$ errors 
from on-shell matrix elements, with $\wt{{\cal O}}^{\Gamma}_{\rm R}$ 
a dimension 4 operator with derivative.
In a previous paper \cite{improved-Z} we have evaluated $Z_\Gamma$
for a class of improved gluon actions.
Our task now is to generalize the analysis to (i) check that there are 
no ${\cal O}(g^2 a\log a)$ errors with the 
operator (\ref{eqn:improved-onshell}), and (ii) determine
the ${\cal O}(g^2 a)$ coefficients $B_\Gamma$ and $C_\Gamma$. 
In the following we set the Wilson parameter $r=1$. 

\reseteqnum
\section{Analysis of one-loop amplitudes}

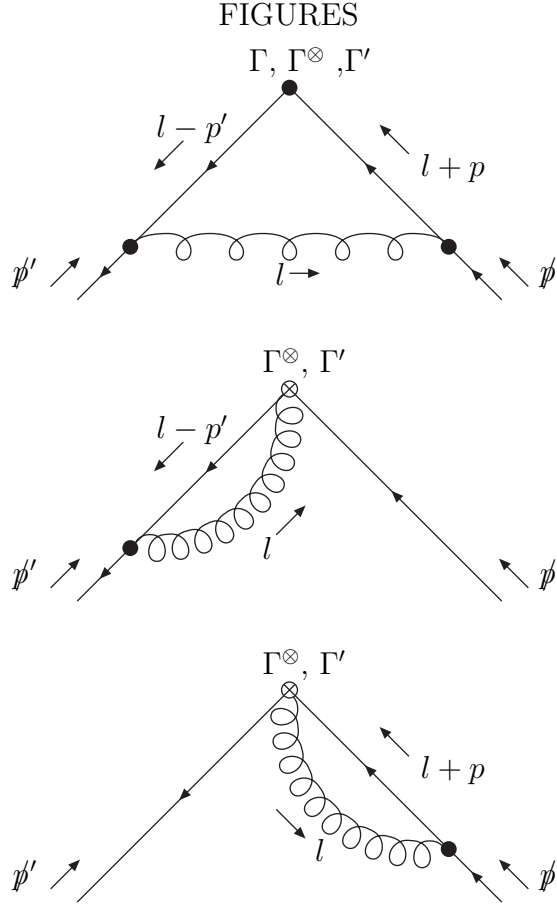
\begin{figure}

\begin{center}\begin{picture}(200,100)(0,0)
\ArrowLine(180,10)(160,30)
\ArrowLine(160,30)(100,90)
\ArrowLine(100,90)(40,30)
\ArrowLine(40,30)(20,10)
\Gluon(40,30)(160,30){5}{5}
\Vertex(40,30){3}
\Vertex(160,30){3}
\Vertex(100,90){3}
\Text(150,60)[l]{$l+p$}\LongArrow(145,65)(135,75)
\Text(50,75)[l]{$l-p'$}\LongArrow(60,70)(50,60)
\Text(95,20)[l]{$l$}\LongArrow(100,20)(110,20)
\Text(85,100)[l]{$\Gamma$, $\Gamma^\otimes$ ,$\Gamma'$}
\Text(195,20)[l]{$\pslash$}\LongArrow(190,15)(180,25)
\Text(5,20)[r]{$\pslash'$}\LongArrow(10,15)(20,25)
\end{picture}
\end{center}
\begin{center}
\setcaption{figure}
\begin{picture}(200,100)(0,0)
\ArrowLine(180,10)(100,90)
\ArrowLine(100,90)(40,30)
\ArrowLine(40,30)(20,10)
\GlueArc(50,80)(50,-100,10){-5}{10}
\Vertex(40,30){3}
\BCirc(100,90){3}
\Line(98,88)(102,92)
\Line(98,92)(102,88)
\Text(90,30)[l]{$l$}\LongArrow(95,35)(105,45)
\Text(50,75)[l]{$l-p'$}\LongArrow(60,70)(50,60)
\Text(90,100)[l]{$\Gamma^\otimes$, $\Gamma'$}
\Text(195,20)[l]{$\pslash$}\LongArrow(190,15)(180,25)
\Text(5,20)[r]{$\pslash'$}\LongArrow(10,15)(20,25)
\end{picture}
\end{center}
\begin{center}
\setcaption{figure}
\begin{picture}(200,100)(0,0)
\ArrowLine(180,10)(160,30)
\ArrowLine(160,30)(100,90)
\ArrowLine(100,90)(20,10)
\GlueArc(150,80)(52,170,280){-5}{10}
\Vertex(160,30){3}
\BCirc(100,90){3}
\Line(98,88)(102,92)
\Line(98,92)(102,88)
\Text(150,60)[l]{$l+p$}\LongArrow(145,65)(135,75)
\Text(110,30)[l]{$l$}\LongArrow(95,45)(105,35)
\Text(90,100)[l]{$\Gamma^\otimes$, $\Gamma'$}
\Text(195,20)[l]{$\pslash$}\LongArrow(190,15)(180,25)
\Text(5,20)[r]{$\pslash'$}\LongArrow(10,15)(20,25)
\end{picture}
\end{center}
\caption{Structure of one-loop diagrams for bilinear quark 
operator (\protect{\ref{eqn:improved-onshell}})}
\label{fig:diagram}
\end{figure}

The structure of one-loop diagrams relevant for our analysis 
is depicted in Fig.~\ref{fig:diagram} where we indicate our 
momentum assignment.  We calculate the corresponding amplitudes 
in Feynman gauge 
imposing the on-shell condition to external momenta,
which becomes for these diagrams,
\begin{eqnarray}
(-i \pslash' + m_R) \Gamma
 = \Gamma ( i \pslash + m_R) =0. 
\label{eqn:onshell}
\end{eqnarray}
We identify the renormalized mass $m_R$, 
to be explicitly defined  below, with the on-shell mass
as there appear no divergences in these diagrams
which should be renormalized into quark mass.
We note that the bare quark mass $m_0$ enters in the field
rotation \eqn{eqn:rotate-l} of the bare operator, where we make use 
of the tree-level equation of motion.

The vertex function in momentum space calculated to one-loop order 
has the form 
\begin{eqnarray}
G^\Gamma &=&
 \left[ 1+a m_0 \left( 1-z \right) \right] \Gamma
 + z a \frac{1}{2} \left( i\pslash' \Gamma - \Gamma i\pslash \right)
\nn\\
&&
 + g^2 C_F \left[ 1+a m_0 \left( 1-z \right) \right] T_\Gamma
 + g^2 C_F z T_{\Gamma^\otimes} - g^2 C_F z^2 T_{\Gamma'} ,
\label{eq:vertex}
\end{eqnarray}
where $\Gamma$ represents the tree level contribution,
$T_\Gamma$, $T_{\Gamma^\otimes}$ and $T_{\Gamma'}$
are one-loop contributions from the vertices 
$\Gamma$, $\Gamma^\otimes$ and $\Gamma '$.

A problem in extracting ${\cal O}(a)$ terms of the one-loop contributions
is that they are infra-red divergent for on-shell external momenta.
We treat this problem by supplying a mass $\lambda$ to the gluon propagator.
The one-loop amplitudes, being functions of $pa, p^\prime a, m_Ra$ and
$\lambda a$, are then finite, which we expand around $pa=p^\prime a=0$
and $m_Ra=0$, keeping $\lambda a$ finite.  For this procedure to be justified,
infra-red divergences which remain in the vertex function $G_\Gamma$
after wave function renormalization should coincide with those in the
continuum.  We check this point explicitly below.

As a first step to extract ${\cal O}(a)$ terms, 
we expand one-loop contributions in terms of external momenta.  Under the 
on-shell condition (\ref{eqn:onshell}) the ${\cal O}(a)$ term can be 
written in two alternative forms, {\it i.e., }
\begin{eqnarray}
T_{(\Gamma, \Gamma^\otimes, \Gamma ')} &=&
V^+_{(\Gamma, \Gamma^\otimes, \Gamma ')} \Gamma
+ v^+_{(\Gamma, \Gamma^\otimes, \Gamma ')}
 i a q_\mu^+ \wt{\Gamma}_\mu^+,\\
&=&
V^-_{(\Gamma, \Gamma^\otimes, \Gamma ')} \Gamma
+ v^-_{(\Gamma, \Gamma^\otimes, \Gamma ')}
 i a q_\mu^- \wt{\Gamma}_\mu^-,
\label{eq:momentumexpansion}
\end{eqnarray}
where $q_\mu^\pm = p_\mu ' \pm p_\mu$, and 
$q_\mu^\pm \wt{\Gamma}_\mu^\pm$ is a dimension-4 operator vertex
with the same quantum number as $\Gamma$ as listed 
in Table~\ref{tbl:gamma-tilda} for each $\Gamma$.
The on-shell identities which relate the two forms are given by 
\begin{eqnarray}
&&
-i q_\mu^+ \gamma_5 
+i q_\nu^- \sigma_{\mu \nu} \gamma_5  + 2 m_R \gamma_\mu \gamma_5 = 0 ,
\\
&&
-i\sigma_{\mu \nu} q_\nu^+ +i q_\mu^- - 2 m_R \gamma_\mu = 0 ,
\\
&&
q_\mu^- \gamma_\mu \gamma_5 = q_\mu^+ \gamma_\mu =0 ,
\\
&&
q_\mu^+ \gamma_\mu \gamma_5 + 2i m_R \gamma_5 = 0 ,
\\
&&
q_\mu^- \gamma_\mu + 2i m_R = 0 ,
\\
&&
 p_\rho' (\sigma_{\rho \mu} \gamma_\nu - \sigma_{\rho \nu} \gamma_\mu) 
-p_\rho  (\gamma_\mu \sigma_{\nu \rho} - \gamma_\nu \sigma_{\mu \rho})
+ (q_\mu^+ \gamma_\nu - q_\nu^+ \gamma_\mu)
 + 4i m_R \sigma_{\mu \nu} = 0.
\end{eqnarray}
We choose to work with $q_\mu^+ \wt{\Gamma}_\mu^+$ and drop the $+$ suffix. 
The momentum $q_\mu^+=q_\mu$ represents the momentum transfer at 
the operator vertex.
We observe from the identities above and Table~\ref{tbl:gamma-tilda} 
that $q_\mu^+ \wt{\Gamma}_\mu^+$ actually vanishes 
by the on-shell condition for the scalar and pseudo scalar operators.
Substituting the expansion (\ref{eq:momentumexpansion}) into 
(\ref{eq:vertex}), we obtain 
\begin{eqnarray}
G^\Gamma 
&=&
\left[ 1+a m_0 \left( 1-z \right) + z a m_R \right] \Gamma
+ g^2 C_F \left[ 1+a m_0 \left( 1-z \right) \right]
 (V_\Gamma \Gamma + v_\Gamma ia q_\mu \wt{\Gamma}_\mu)
\nn\\
&&
+ g^2 C_F z (V_{\Gamma^\otimes} \Gamma
       + v_{\Gamma^\otimes} ia q_\mu \wt{\Gamma}_\mu)
- g^2 C_F z^2 (V_{\Gamma'} \Gamma + v_{\Gamma'} ia q_\mu \wt{\Gamma}_\mu).
\end{eqnarray}

As a second step, we expand 
the one-loop amplitudes $V_{(\Gamma, \Gamma^\otimes, \Gamma ')}$ 
in terms of the lattice spacing $a$ multiplied by the renormalized
quark mass $m_R$.  This leads to
\begin{eqnarray}
&&
V_\Gamma = \frac{h_2(\Gamma)}{4} L
 + V_\Gamma^{(0)} + am_R V_\Gamma^{(1)} + {\cal O}(a^2) , 
\\
&&
V_{\Gamma^\otimes} =  h_2' (\Gamma) am_R L
 + V_{\Gamma^\otimes}^{(0)} + am_R V_{\Gamma^\otimes}^{(1)} +{\cal O}(a^2) ,
\\
&&
V_\Gamma' = V_{\Gamma'}^{(0)} + am_R V_{\Gamma'}^{(1)} + {\cal O}(a^2) ,
\end{eqnarray}
where $V_{(\Gamma, \Gamma^\otimes, \Gamma ')}^{(0,1)}$
are constants independent of $a$ and $g$.
In a similar expansion of $v_{(\Gamma, \Gamma^\otimes, \Gamma ')}$, 
we only need to keep the leading term in $a$, 
and hence they can be regarded as 
a constant as well.  
The logarithmic divergence $L$ is defined as
\[L = -\frac{1}{16\pi^2} \log \lambda^2 a^2
 \sim \int_{-\pi}^\pi \frac{d^4 l}{(2\pi)^4}
\frac{1}{l^2} \frac{1}{l^2 + \lambda^2}a^2\]
with $\lambda$ being gluon mass to regularize the infra-red divergence 
which appears in on-shell vertex functions.
The coefficients $h_2 (\Gamma)$ and $h_2' (\Gamma)$ are
given in Table~\ref{tbl:h2}.

In order to relate bare operators to renormalized operators 
we further need the renormalization factor for quark wave function $Z_\psi$
and mass $Z_m$, which are defined as  
\begin{eqnarray}
\psi_0 &=& Z_\psi^{-1/2} \psi_R ,
\\
m_0 &=& Z_m^{-1} m_R + g^2 C_F \frac{\Sigma_0}{a}.
\end{eqnarray}
The explicit form of these factors are obtained from the inverse full quark 
propagator expanded to ${\cal O}(g^2 a)$,
\begin{eqnarray}
S_F^{-1} &=& i\pslash +m_0 + \frac{1}{2}a p^2 -\Sigma (p, m_0),
\end{eqnarray}
where the one-loop correction to self energy can be written as 
\begin{eqnarray}
\Sigma (p, m_0) &=&
g^2 C_F \Biggl[ \frac{\Sigma_0}{a}
+i \pslash \left( -L +\Sigma_1 \right)
+m_0 \left( -4L +\Sigma_2 \right)
\nn\\
&&
+ap^2 \left( -L+\sigma_1 \right)
+am_0 i\pslash \left( 4L + \sigma_2 \right)
+am_0^2 \left( 2L + \sigma_3 \right)
\Biggl] + {\cal O}(a^2)
\end{eqnarray}
with $\Sigma_{0, 1, 2}$ and $\sigma_{1,2,3}$ being constants. 
In terms of these constants, the renormalization factors are given by 
\begin{eqnarray}
Z_\psi^{-1} &=& 1 + g^2 C_F \,\left( - L +{\Sigma_1} \right)
 + a\,m \left( -1
 + \,g^2 C_F \, \left(L
 + 2\,{\sigma_1} + {\sigma_2} - 3\,{\Sigma_1} + {\Sigma_2}
\right) \right) ,
\\
Z_m^{-1} &=& 1 + \,g^2 C_F \left( -3\,L - {\Sigma_1} + {\Sigma_2} \right)
 + \,a m  \left( {\frac{1}{2}}
 + \frac{g^2 C_F}{2} \left(
 -3\,L - 2\,{\sigma_1} - 2\,{\sigma_2} + 2\,{\sigma_3} + {\Sigma_1}
 \right) \right) .
\end{eqnarray}
Here we have defined 
\begin{equation}
m=m_0-g^2 C_F \frac{\Sigma_0}{a},  
\label{eq:submass}
\end{equation}
and have made use of the relation $g^2 m_0 = g^2 m + {\cal O}(g^4)$.
The wave function renormalization factor can be rewritten 
in terms of the renormalized mass,
\begin{eqnarray}
Z_\psi^{-1} =
1 + g^2 C_F \left( -L + \Sigma_1 \right)
  + a m_R \left( -1 
 + g^2 C_F \left( 4  L - z_m + \Sigma_1^{(1)} \right) \right),
\label{eqn:Z2}
\end{eqnarray}
where
\begin{eqnarray}
&&z_m = - {\Sigma_1} + {\Sigma_2} ,
\\
&&\Sigma_1^{(1)} = 2\,{\sigma_1}+{\sigma_2}-3\,{\Sigma_1}+{\Sigma_2} .
\end{eqnarray}

Replacing the quark mass $m_0$ with the renormalized mass $m_R$,  
we obtain for the the vertex,
\begin{eqnarray}
G^\Gamma
&=&
 \left[ 1+a m_R +
 g^2 C_F L \left( \frac{h_2(\Gamma)}{4}
 + am_R \left( \frac{h_2(\Gamma)}{4} (1-z)
 + h_2' (\Gamma) z -3 (1-z) \right) \right) \right] \Gamma
\nn\\
&&
+ g^2 C_F \left[
\Sigma_0 (1-z) + V_\Gamma^{(0)} +z V_{\Gamma^\otimes}^{(0)} 
-z^2 V_{\Gamma'}^{(0)}
 \right] \Gamma
\nn\\
&&
+ g^2 C_F a m_R \left[
 z_m (1-z) + (1-z) V_\Gamma^{(0)} + V_\Gamma^{(1)}
+ z V_{\Gamma^\otimes}^{(1)} - z^2 V_{\Gamma'}^{(1)}
\right] \Gamma
\nn\\
&&
+ g^2 C_F (v_\Gamma + z v_{\Gamma^\otimes} -z^2  v_{\Gamma'} )
 i a q_\mu \wt{\Gamma}_\mu
\nn\\
&=&
 \left[ 1+a m_R +
 g^2 C_F L \left( \frac{h_2(\Gamma)}{4}
 + am_R \left( \frac{h_2(\Gamma)}{4} -3 \right) \right) \right] \Gamma
\nn\\
&&
+ g^2 C_F \left[
\Sigma_0 (1-z) + V_\Gamma^{(0)} +z V_{\Gamma^\otimes}^{(0)} 
-z^2 V_{\Gamma'}^{(0)}
 \right] \Gamma
\nn\\
&&
+ g^2 C_F a m_R \left[
 z_m (1-z) + (1-z) V_\Gamma^{(0)} + V_\Gamma^{(1)}
+ z V_{\Gamma^\otimes}^{(1)} - z^2 V_{\Gamma'}^{(1)}
\right] \Gamma
\nn\\
&&
+ g^2 C_F (v_\Gamma + z v_{\Gamma^\otimes} -z^2  v_{\Gamma'} )
 i a q_\mu \wt{\Gamma}_\mu ,
\label{eqn:effective-vertex}
\end{eqnarray}
where the relation
$h_2(\Gamma)/4 - h_2'(\Gamma) = 3$, valid for each $\Gamma$,  
is used for the second equality.

We now multiply the vertex function $G^\Gamma$ by the quark wave function 
renormalization factor $Z_\psi^{-1}$ of \eqn{eqn:Z2}.  
The ${\cal O}(a)$ and ${\cal O}(g^2 a\log a)$ terms all cancel out in the 
combination $Z_\psi^{-1} G^\Gamma$ for arbitrary values of the parameter $z$, 
and the result can be written as an operator identity 
\eqn{eqn:renormalization} with the constants given by 
\begin{eqnarray}
Z_\Gamma^{-1} &=&
1 + g^2 C_F \left( \left( \frac{h_2(\Gamma)}{4}-1 \right)L+
\Sigma_1 + V_\Gamma^{(0)} + \Sigma_0 (1-z) + z V_{\Gamma^\otimes}^{(0)}
-z^2 V_{\Gamma'}^{(0)}
 \right) ,
\label{eq:zfactor}
\\
B_\Gamma &=& -\Biggl(
 \Sigma_1 +\Sigma_1^{(1)} + V_\Gamma^{(1)} - \Sigma_0 (1-z)
+z (-z_m-V_\Gamma^{(0)}-V_{\Gamma^\otimes}^{(0)}+V_{\Gamma^\otimes}^{(1)}) 
\nn\\
&&
 +z^2 \left( V_{\Gamma'}^{(0)} - V_{\Gamma'}^{(1)} \right) \Biggr) ,
\label{eq:bgamma}
\\
C_\Gamma &=&
-\left( v_\Gamma + z v_{\Gamma^\otimes} -z^2  v_{\Gamma'} \right).
\label{eq:cgamma}
\end{eqnarray}

Let us add a remark on infra-red divergence.
For $v_{(\Gamma, \Gamma^\otimes, \Gamma')}$ that enter in the coefficient 
$C_\Gamma$, these divergences cancel out in the total contribution 
at each order of $z$.  On the other hand, infra-red divergences remain in 
$V_{(\Gamma, \Gamma^\otimes, \Gamma')}$. 
The divergence in the ${\cal O}(g^2a)$ term in $G^\Gamma$ are, however, 
canceled by that of the quark wave function renormalization factor.
The remaining divergence, which is of form $g^2L$ and appears in 
$Z_\Gamma$, coincides with that which is present in on-shell vertex
functions for the renormalized operator in the continuum. 

In the calculation above we employed the bare operator
\eqn{eqn:improved-onshell} which contains the bare quark mass $m_0$.
It is possible to replace $m_0$ by the subtracted mass $m$ of 
(\ref{eq:submass}).  Defining
\begin{eqnarray}
{\cal O}_0^\Gamma
= \left[ 1+ar \left( 1-z \right) m \right] \bpsi \Gamma \psi
+ z \Gamma^\otimes
- z^2 \Gamma ',
\label{eq:bilinearsubtracted}
\end{eqnarray}
it is straightforward to check that ${\cal O}(a)$ and 
${\cal O}(g^2 a \log a)$ terms also cancel for this operator; 
$\Sigma_0$ plays no role in improving the operator. 
The renormalization coefficients for this operator, which is more 
convenient for practical use, are obtained from 
(\ref{eq:zfactor}-\ref{eq:cgamma}) by eliminating $\Sigma_0$. 

\reseteqnum
\section{Results for the one-loop coefficients}

Manipulations in the previous section have reduced the determination of 
one-loop coefficients to an evaluation of a number of integral constants.  
Working out the integrands for the integrals is 
a straightforward but tedious task, which we carry out by 
{\it Mathematica}. 
The output is converted to a FORTRAN code, and the integrals are 
evaluated by the Monte Carlo routine VEGAS in double precision. 
We employ $20$ sets of $10^5$ points for integration except for 
$C_A$ for $z=0$ for which we use 20 sets of $10^6$ points. 
Errors are estimated from variation of integrated
values over the sets.  

For the gluon action we consider the form given by 
\begin{equation}
S_{\rm gluon} = \frac{1}{g^2}\left\{
c_0 \sum_{plaquette} {\rm Tr}U_{pl}
+ c_1  \sum_{rectangle} {\rm Tr} U_{rtg}
+ c_2 \sum_{chair} {\rm Tr} U_{chr}
+ c_3 \sum_{parallelogram} {\rm Tr} U_{plg}\right\},
\end{equation}
where the first term represents the standard plaquette term, and the
remaining terms are six-link loops formed by a $1\times 2$ rectangle,
a bent $1\times 2$ rectangle (chair) and a 3-dimensional parallelogram.
The coefficients $c_0, \cdots, c_3$ satisfy the normalization condition
\begin{equation}
c_0+8c_1+16c_2+8c_3=1.
\end{equation}

At the one-loop level, the choice of the gluon action is specified by the
pair of numbers $c_1$ and $c_{23}=c_2+c_3$.  We calculate the constants 
for five cases: 
(i) the standard plaquette action $c_1=0, c_{23}=0$, 
(ii) the tree-level improved action in the Symanzik approach
$c_1=-1/12, c_{23}=0$\cite{Weisz83},
and (iii) three choices suggested
by an approximate renormalization-group analysis, $c_1=-0.331, c_{23}=0$
and $c_1=-0.27, c_{23}=-0.04$ by Iwasaki\cite{Iwasaki83},
and $c_1=-0.252, c_{23}=-0.17$ by Wilson\cite{Wilson80}.

Let us write ${\cal O}_{\overline{MS}}^{\Gamma}$ for the renormalized 
bilinear operator in the continuum in the $\ovl{\rm MS}$ scheme, and 
define
\begin{equation}
{\cal O}_{\overline{MS}}^{\Gamma}=Z_{\overline{MS}}^\Gamma {\cal O}_0^\Gamma ,
\end{equation}
where ${\cal O}_0^\Gamma$ denotes the lattice bare operator and 
\begin{equation}
Z_{\overline{MS}}^\Gamma=1+\frac{g^2 C_F}{16\pi^2} 
\left( \left( \frac{h_2(\Gamma)}{4}-1 \right)\log (\mu a)^2 +
z_\Gamma \right) .
\end{equation}
Results for the finite constant $z_\Gamma$ for the lattice operator 
(\ref{eqn:improved-onshell}) rotated with the bare mass $m_0$ 
have already been given in a previous paper\cite{improved-Z}. 
We list in Table~\ref{tbl:zfact} the values of $z_\Gamma$ for 
the operator (\ref{eq:bilinearsubtracted}) 
defined with the subtracted mass $m$ at $z=0$ for completeness.

Our new results for the one-loop coefficients $C_\Gamma$ are given 
in Table~\ref{tbl:cs}, and those for $B_\Gamma$ in  
Tables \ref{tbl:ba} and \ref{tbl:bp}
for the definition excluding $\Sigma_0$.  Numerical values are given for 
the coefficients of expansion in $z$ defined as 
\begin{eqnarray}
B_\Gamma &=& B_\Gamma^{(0)} + z B_\Gamma^{(1)} - z^2 B_\Gamma^{(2)},
\\
C_\Gamma &=& C_\Gamma^{(0)} + z C_\Gamma^{(1)} - z^2 C_\Gamma^{(2)}.
\end{eqnarray}
In the tensor channel only $C_T^{(0)}$ and $B_T^{(0)}$ are evaluated as 
necessity for the operator in this  channel does not seem 
to warrant a CPU time-consuming calculation of integrands which are more 
complex than the other cases. 
In Tables \ref{tbl:ba} and \ref{tbl:bp} a general trend is apparent 
that the coefficients are reduced by roughly 
a factor two for renormalization-group improved gluon actions as compared to 
those for the plaquette action, as already observed for 
$z_\Gamma$\cite{improved-Z}.

Comparison of our results with those of Refs.~\cite{Luescher9606,Sint-Weisz}
obtained with the Schr\"odinger functional is made in the following way. 
The authors of these references start from a local bilinear operator 
\begin{eqnarray}
{\cal O}_0^\Gamma = \bpsi \Gamma \psi,
\end{eqnarray}
and relate it to the renormalized operator through 
\begin{eqnarray}
{\cal O}^\Gamma_0 = Z^{-1}_\Gamma ( g^2, a ) {\cal O}^\Gamma_R
- c_\Gamma i q_\mu \wt{{\cal O}}^\Gamma_\mu.
\end{eqnarray}
The renormalization factor $Z_\Gamma $ is expanded in the 
lattice spacing $a$ as follows,
\begin{eqnarray}
Z_\Gamma (g^2, a )&=& 
Z_{0\,\Gamma} ( g^2 )\left( 1 + a m b_\Gamma(g^2) \right).
\end{eqnarray}
where $Z_{0\,\Gamma} ( g^2 )$ does not contain terms of ${\cal O}(a)$, 
${\cal O}(g^2a\log a)$ or ${\cal O}(g^2a)$. 
The renormalization factor for the quark field $\psi$ and the quark 
mass $m$ are expanded in a similar manner. 

With these definitions, the vertex function of the bare operator has the 
form
\begin{eqnarray}
G^\Gamma
=
\Gamma
+ g^2 C_F \left(
 \frac{h_2(\Gamma)}{4} L+ V_\Gamma^{(0)} + am V_\Gamma^{(1)}
 \right) \Gamma
+ g^2 C_F a v_\Gamma i q_\mu \wt{\Gamma}_\mu.
\end{eqnarray}
The expressions for the wave function, quark mass 
and quark bilinear operator renormalization factors are given by 
\begin{eqnarray}
Z_\psi^{-1/2} &=&
\left( 1 + \frac{g^2 C_F}{2} \left( -L+ \Sigma_1 \right) \right)
\left( 1 + am \left(-\frac{1}{2}
 +g^2 C_F \left(
{\sigma_1} + \frac{\sigma_2}{2} - \,{\Sigma_1} + \frac{\Sigma_2}{2}
\right) \right) \right) ,
\\
Z_m &=&
\left( 1+ g^2 C_F \left(3 L + \left( \Sigma_1 - \Sigma_2 \right) \right)
 \right)
\left( 1 + am \left( -\frac{1}{2}
 + g^2 C_F \left(
{\sigma_1}+{\sigma_2}-{\sigma_3}-{\Sigma_1}+\frac{1}{2}{\Sigma_2}
\right) \right) \right) ,
\\
Z_\Gamma &=&
\left( 1 - g^2 C_F \left( \left( \frac{h_2(\Gamma)}{4}-1 \right) L
 + \Sigma_1 + V_\Gamma^{(0)} \right) \right)
\left(1 + a m \left( 1-
g^2 C_F \left( \Sigma_1 + \Sigma_1^{(1)} + V_\Gamma^{(1)} \right)
\right) \right).
\end{eqnarray}
Making an expansion 
\begin{eqnarray}
b_x&=&b_x^{(0)}+g^2 C_F b_x^{(1)}+\cdots, \qquad x=\Gamma, \psi, m,\\
c_\Gamma&=&g^2 C_F c_\Gamma^{(1)}+\cdots,
\end{eqnarray} 
we find that the ${\cal O}(g^2 a)$ coefficients are given as
\begin{eqnarray}
b_\psi^{(1)} &=&
 {\sigma_1} + \frac{\sigma_2}{2} - \,{\Sigma_1} + \frac{\Sigma_2}{2} ,
\\
b_m^{(1)} &=&
 {\sigma_1}+{\sigma_2}-{\sigma_3}-{\Sigma_1}+\frac{1}{2}{\Sigma_2} m ,
\\
b_\Gamma^{(1)} &=&
 -\left( \Sigma_1 + \Sigma_1^{(1)} + V_\Gamma^{(1)} \right),
\\
c_\Gamma^{(1)} &=& - v_\Gamma .
\end{eqnarray}

Comparing these expressions with (\ref{eq:bgamma}) and (\ref{eq:cgamma}), 
we see that $c_\Gamma^{(1)}$ equals our 
$C_\Gamma^{(0)}$ for $z=0$ tabulated in Table~\ref{tbl:cs}, 
and $b_\Gamma^{(1)}$ equals our $B_\Gamma^{(0)}$ for $z=0$
without the $\Sigma_0$ term given in Tables \ref{tbl:ba} and
\ref{tbl:bp}.
Our results for $b_\psi^{(1)}$ and $b_m^{(1)}$ are given in table \ref{tbl:bm},
where we also list the contribution from the wave function renormalization 
factor $b_0= -(\Sigma_1 + \Sigma_1^{(1)})=-2 b_\psi^{(1)}$.

In Table~\ref{tbl:comparison} we collect our results for the mixing 
coefficients for $z=0$ for the plaquette gluon action, 
and compare them with those of Refs.~\cite{Luescher9606,Sint-Weisz}. 
As we already remarked, we employ 20 sets of $10^6$ points for evaluating 
$c_A^{(1)}$ by VEGAS in this table, with which we find a complete agreement 
with the result of Refs.~\cite{Luescher9606,Sint-Weisz}.  Good agreement is 
also found for all the other coefficients obtained with 20 sets of $10^5$ 
points.  We do not pursue more precise evaluation for the latter coefficients 
since it would require significantly more computing power due to an increased 
complexity of integrands and the number of terms.

It was observed in Ref.~\cite{Sint-Weisz} that the values of $b_\Gamma^{(1)}$ 
are close to each other.  Numerically this arises from the fact that 
the contribution from the wave function renormalization $b_0$, 
common to various Dirac channels $\Gamma$, dominates
over the vertex contributions.  

\section{Concluding remarks}

In this article we have carried out a perturbative evaluation of vertex 
functions to determine the $O(g^2a)$ mixing coefficients of bilinear 
quark operators.  For the standard plaquette action for gluons, our results
agree with those obtained previously with the Schr\"odinger functional 
method.
We have also generalized the determination to a class of improved gluon 
actions for use in numerical simulations employing such actions. 

Our calculations are carried out by an expansion of vertex functions 
regarding external momenta and renormalized quark mass $m_R$ as small in units
of lattice spacing $a$.   Hence the present work does not cover the case of 
heavy quark such that $m_Ra>O(1)$.
It has been pointed out recently 
in a one-loop calculation in NRQCD\cite{Shigemitsu-Morningstar} that    
the mixing coefficient $c_A^{(1)}$ for heavy-light axial vector current 
is large compared to the value for the light-light case treated here.
In our calculation a significant cancellation is observed between 
terms from various diagrams contributing to $c_A^{(1)}$.    
To understand whether the large value of $c_A^{(1)}$ for heavy quarks 
results from lifting of such a cancellation requires an extension of 
our calculation without making an expansion in 
$m_Ra$\cite{FNAL,Kuramashi,SHIT}.

\section*{Acknowledgements}

We thank Sinya Aoki for informative correspondence. 
Numerical calculations for the present work have been carried
out at the Center for Computational Physics, University of Tsukuba, 
and at Research Institute for Fundamental Physics, Kyoto University.
This work is supported in part by the Grants-in-Aid for
Scientific Research of the Ministry of Education, Science and Culture
(Nos. 2373, 09304029).
Y. T. is supported by Japan Society for Promotion of Science.

\begin{table}[bht]
\caption{Dimension 4 vertex mixing with
 $\bpsi \Gamma \psi$ at one loop order.}
\label{tbl:gamma-tilda}
\begin{center}
\begin{tabular}{l|ll}
$\Gamma$              & $q_\mu^+ \wt{\Gamma}_\mu^+$
 & $q_\mu^- \wt{\Gamma}_\mu^-$ \\
\hline 
$\gamma_\mu \gamma_5$ & $q_\mu^+ \gamma_5$
 & $q_\nu^- \sigma_{\mu \nu} \gamma_5$ \\
$\gamma_\mu$          & $q_\nu^+ \sigma_{\mu \nu}$
 & $q_\mu^-$ \\
$\gamma_5$            & $q_\mu^+ \gamma_\mu \gamma_5$
 & $q_\mu^- \gamma_\mu \gamma_5$ \\
$1$                   & $q_\mu^+ \gamma_\mu$
 & $q_\mu^- \gamma_\mu$ \\
$\sigma_{\mu \nu}$    & $q_\mu^+ \gamma_\nu - q_\nu^+ \gamma_\mu$
 & $p_\rho' (\sigma_{\rho \mu} \gamma_\nu - \sigma_{\rho \nu} \gamma_\mu) 
   -p_\rho (\gamma_\mu \sigma_{\nu \rho} - \gamma_\nu \sigma_{\mu \rho})$ 
\\
\end{tabular}
\end{center}
\end{table}%

\begin{table}[bht]
\caption{Coefficients of logarithmically divergent term.}
\label{tbl:h2}
\begin{center}
\begin{tabular}{l|ll}
$\Gamma$              & $h_2 (\Gamma)$ & $h_2' (\Gamma)$ \\
\hline 
$\gamma_\mu \gamma_5$ & $4$            & $-2$ \\
$\gamma_\mu$          & $4$            & $-2$ \\
$\gamma_5$            & $16$           & $1$  \\
$1$                   & $16$           & $1$  \\
$\sigma_{\mu \nu}$    & $0$            & $-3$ \\
\end{tabular}
\end{center}
\end{table}%

\begin{table}[bht]
\caption{Finite part $z_\Gamma$ of renormalization factor for 
bilinear quark operators ($z=0$).
Coefficients of the term $c_{SW}^n (n=0,1,2)$ in an expansion 
$z_\Gamma=z_\Gamma^{(0)}+c_{SW}z_\Gamma^{(1)}+c_{SW}^2z_\Gamma^{(2)}$ 
are given in the column labeled $(n)$. Errors are at most in the last 
digit given.}
\label{tbl:zfact}
\begin{center}
\begin{tabular}{ll|lll|lll}
\multicolumn{2}{c|}{gauge action}&
\multicolumn{3}{c|}{$V$} &
\multicolumn{3}{c}{$A$} \\
$c_1$ & $c_{23}$ & 
(0) & (1) & (2)  &
(0) & (1) & (2)  \\
\hline
  0    &0    &-20.618&4.745&0.543&-15.797&-0.248&2.251\\
 -1/12 &0    &-16.603&4.228&0.464&-12.540&-0.198&2.021\\
 -0.331&0    &-11.099&3.326&0.336& -8.192&-0.125&1.610\\
 -0.27 &-0.04&-11.540&3.418&0.353& -8.523&-0.131&1.657\\
 -0.252&-0.17&-10.525&3.248&0.338& -7.707&-0.117&1.587\\
\end{tabular}

\begin{tabular}{ll|lll|lll|lll}
\multicolumn{2}{c|}{gauge action}&
\multicolumn{3}{c|}{$S$} &
\multicolumn{3}{c|}{$P$} & 
\multicolumn{3}{c}{$T$} \\
$c_1$ & $c_{23}$ & 
(0) & (1) & (2)  &
(0) & (1) & (2)  &
(0) & (1) & (2)  \\
\hline
 0    &0    &-12.953&-7.738&1.380&-22.596&2.249&-2.036&-17.018&3.913&1.972\\
-1/12 &0    & -9.607&-6.836&1.367&-17.734&2.015&-1.745&-13.539&3.490&1.719\\
-0.331&0    & -4.858&-5.301&1.266&-10.673&1.601&-1.281& -8.939&2.751&1.300\\
-0.27 &-0.04& -5.260&-5.454&1.292&-11.292&1.644&-1.316& -9.283&2.827&1.344\\
-0.252&-0.17& -4.366&-5.166&1.287&-10.001&1.565&-1.212& -8.427&2.687&1.271\\
\end{tabular}
\end{center}
\end{table}

\begin{table}[bht]
\caption{
Mixing coefficients $C_A$, $C_V$, $C_T$
for axial vector, vector and tensor currents.
Coefficients of the term $z^n (n=0,1,2)$ are given in the column 
marked as $(n)$.
$C_T^{(1)}$ and $C_T^{(2)}$ are not calculated.
}
\label{tbl:cs}
\begin{center}
\begin{tabular}{ll|lll|lll|l}
\multicolumn{2}{c|}{gauge action}&
\multicolumn{3}{c|}{$C_A$} &
\multicolumn{3}{c|}{$C_V$} & 
\multicolumn{1}{c}{$C_T$} \\
$c_1$ & $c_{23}$ & 
(0) & (1) & (2)  &
(0) & (1) & (2)  &
(0) \\
\hline
0      & 0     & -0.005680(2)&-0.002316(3)&-0.003808(1) &
    -0.01226(3)&-0.01560(4)&-0.000217(9) &-0.00898(1) \\
-1/12  & 0     & -0.00451(1)&-0.00179(2)&-0.002781(7) &
    -0.01030(4)&-0.01262(4)&0.000034(8) &-0.00741(1)  \\
-0.331 & 0     & -0.00285(1)&-0.00099(2)&-0.001579(6) &
    -0.00729(4)&-0.00825(4)&0.000175(7) &-0.00508(1)  \\
-0.27  & -0.04 & -0.00302(1)&-0.00108(2)&-0.001660(6) &
    -0.00757(4)&-0.00858(4)&0.000159(7) &-0.00530(1)  \\
-0.252 & -0.17 & -0.00281(1)&-0.00098(2)&-0.001431(6) &
    -0.00705(4)&-0.00772(4)&0.000153(7) &-0.00495(1)  \\
\end{tabular}
\end{center}
\end{table}

\begin{table}[bht]
\caption{
Mixing coefficients $B_A$, $B_V$, $B_T$
for axial vector, vector and tensor currents.
Coefficients of the term $z^n (n=0,1,2)$ are given in the column 
marked as $(n)$.
$B_T^{(1)}$ and $B_T^{(2)}$ are not calculated.
}
\label{tbl:ba}
\begin{center}
\begin{tabular}{ll|lll|lll|l}
\multicolumn{2}{c|}{gauge action}&
\multicolumn{3}{c|}{$B_A$} &
\multicolumn{3}{c|}{$B_V$} &
\multicolumn{1}{c}{$B_T$} \\
$c_1$ & $c_{23}$ & 
(0) & (1) & (2)  &
(0) & (1) & (2)  &
(0)  \\
\hline
0      & 0     &
0.1141(1)&-0.0846(1)&0.01637(3)& 0.1150(2)&-0.0442(2)&0.03255(5)& 0.1044(1) \\
-1/12  & 0     &
0.0881(1)&-0.0666(1)&0.01328(3)& 0.0886(2)&-0.0353(2)&0.02551(4)& 0.0795(1) \\
-0.331 & 0     &
0.0547(1)&-0.0419(1)&0.00867(3)& 0.0550(2)&-0.0228(1)&0.01583(4)& 0.0482(1) \\
-0.27  & -0.04 &
0.0572(1)&-0.0438(1)&0.00909(3)& 0.0575(2)&-0.0238(1)&0.01656(4)& 0.0505(1) \\
-0.252 & -0.17 &
0.0512(1)&-0.0393(1)&0.00827(3)& 0.0514(2)&-0.0213(1)&0.01479(4)& 0.0448(1) \\
\end{tabular}
\end{center}
\end{table}

\begin{table}[bht]
\caption{Mixing coefficients $B_P$, $B_S$ for pseudo scalar and  
scalar density.
Coefficients of the term $z^n (n=0,1,2)$ are given in the column 
marked as $(n)$.
}
\label{tbl:bp}
\begin{center}
\begin{tabular}{ll|lll|lll}
\multicolumn{2}{c|}{gauge action}&
\multicolumn{3}{c|}{$B_P$} &
\multicolumn{3}{c}{$B_S$} \\
$c_1$ & $c_{23}$ & 
(0) & (1) & (2)  &
(0) & (1) & (2)  \\
\hline
0      & 0     &
  0.1148(1)&-0.01698(6)&0.03788(5) & 0.1444(2)&-0.0535(2)&0.00864(6) \\
-1/12  & 0     &
  0.0890(1)&-0.01378(6)&0.02872(4) & 0.1144(2)&-0.0403(2)&0.00718(5) \\
-0.331 & 0     &
  0.0561(1)&-0.00909(6)&0.01697(4) & 0.0747(2)&-0.0235(1)&0.00475(4) \\
-0.27  & -0.04 &
  0.0586(1)&-0.00937(6)&0.01782(4) & 0.0777(2)&-0.0248(2)&0.00501(4) \\
-0.252 & -0.17 &
  0.0527(5)&-0.00816(6)&0.01569(4) & 0.0706(2)&-0.0221(2)&0.00452(4) \\
\end{tabular}
\end{center}
\end{table}

\begin{table}[bht]
\caption{Mixing coefficients $b_\psi^{(1)}$, $b_m^{(1)}$
for quark operator. Values for $b_0$ are also included. }
\label{tbl:bm}
\begin{center}
\begin{tabular}{ll|lll}
\multicolumn{2}{c|}{gauge action}& & \\
$c_1$ & $c_{23}$ & 
$b_\psi^{(1)}$ & $b_m^{(1)}$ & $b_0$ \\
\hline
0      & 0     &  -0.05191(3)&-0.07218(5)&0.10381(9)\\
-1/12  & 0     &  -0.03968(3)&-0.05722(5)&0.07937(9)\\
-0.331 & 0     &  -0.02430(3)&-0.03737(5)&0.04860(9)\\
-0.27  & -0.04 &  -0.02543(3)&-0.03891(5)&0.05086(9)\\
-0.252 & -0.17 &  -0.02262(3)&-0.03526(5)&0.04525(9)\\
\end{tabular}
\end{center}
\end{table}

\begin{table}[bht]
\caption{Comparison with previous work\protect\cite{Luescher9606,Sint-Weisz}
for the plaquette gluon action.}
\label{tbl:comparison}
\begin{center}
\begin{tabular}{l|llllll}
 & $c_A^{(1)}$ & $c_V^{(1)}$ & $b_m^{(1)}$ & $b_A^{(1)}$ & $b_V^{(1)}$ &
 $b_P^{(1)}$ \\
\hline
Ours       & -0.005680(2)&-0.01226(3)& -0.07218(5)&0.1141(1) &0.1150(2) &
0.1148(1) \\
Sint-Weisz\protect\cite{Sint-Weisz} 
           & -0.005680(2)&-0.01225(1)& -0.07217(2)& 0.11414(4)&0.11492(4)&
0.11484(2)\\
\end{tabular}
\end{center}
\end{table}

\end{document}